\documentclass[sigconf]{acmart}
\usepackage{pgfplots}
\pgfplotsset{compat=newest}
\usepackage{graphicx}
\AtBeginDocument{%
  }

\copyrightyear{2025}
\acmYear{2025}
\setcopyright{rightsretained}
\acmConference[SenSys '25]{The 23rd ACM Conference on Embedded Networked Sensor Systems}{May 6--9, 2025}{Irvine, CA, USA}
\acmBooktitle{The 23rd ACM Conference on Embedded Networked Sensor Systems (SenSys '25), May 6--9, 2025, Irvine, CA, USA}
\acmDOI{10.1145/3715014.3724040}
\acmISBN{979-8-4007-1479-5/2025/05}

\begin{document}

\title{Poster Abstract: Time Attacks using Kernel Vulnerabilities}

\author{Muhammad Abdullah Soomro}
\affiliation{%
  \institution{University of Massachusetts, Amherst}
  \country{USA}}
\email{msoomro@umass.edu}

\author{Adeel Nasrullah}
\affiliation{%
  \institution{University of Massachusetts, Amherst}
  \country{USA}
}
\email{anasrullah@umass.edu}

\author{Fatima Muhammad Anwar}
\affiliation{%
  \institution{University of Massachusetts, Amherst}
  \country{USA}
}
\email{fanwar@umass.edu}

\renewcommand{\shortauthors}{Soomro et al.}

\begin{abstract}
Timekeeping is a fundamental component of modern computing; however, the security of system time remains an overlooked attack surface, leaving critical systems vulnerable to manipulation. This paper examines time manipulation attacks that exploit kernel vulnerabilities to distort an application’s perception of time. We categorize these attacks into constant, incremental, and randomized delay strategies and analyze their impact on system performance. Through experimental evaluation, we demonstrate how adversaries can manipulate system time via dynamic library injection and syscall modification, disrupting time-sensitive applications. While Trusted Execution Environments (TEEs) offer partial isolation, they fail to address time security concerns fully. Our results highlight the challenges of securing system time and underscore the need for robust mitigation strategies.
\end{abstract}



\maketitle

\section{Introduction}
Accurate timekeeping is fundamental to modern computing and critical for cryptographic protocols, distributed systems, and real-time infrastructure operations. However, system time security often remains overlooked, creating vulnerabilities. Prior research demonstrates that timing attacks can disrupt sensor networks or disable functionalities in smart power grids~\cite{Time-sync-attacks, zhang2013time}. Kernel vulnerabilities can thus enable attackers to compromise time-sensitive processes, impacting system functionality and data integrity.

Timekeeping spans hardware timers, kernel subsystems, and user-level software. Attackers with kernel-level privileges may manipulate system calls or hardware timers, affecting distributed databases, blockchain protocols, and autonomous systems through delay attacks, time spoofing, or system call rerouting~\cite{timeseal,timeguard}.

Current defenses primarily utilize trusted execution environments (TEEs) like Intel SGX or ARM TrustZone to isolate critical computations~\cite{juffinger2024like}. However, TEEs have significant limitations: they impose high overhead, restrict user-level application access, and depend on untrusted external synchronization. Moreover, TEEs struggle with maintaining monotonicity when adversaries control enclave scheduling~\cite{triad}.

We focus on two primary attack vectors: privilege escalation and code injection. Privilege escalation facilitates kernel manipulations, including dynamic library injection and system call modifications. Code injection targets user-space applications by embedding malicious instructions into timekeeping logic. We experimentally quantify these attacks' performance impact using benchmarks such as \texttt{sysbench} and \texttt{rt-bench}~\cite{rt-bench}.

\vspace{-0.3cm}
\section{Timing Attacks}
We focus on two prevalent kernel vulnerabilities: privilege escalation and code injection. Privilege escalation attacks exploit weaknesses in system access controls, allowing adversaries to gain elevated permissions and manipulate core system functionalities. In timekeeping, attackers can alter how time-related data is generated, accessed, or communicated to user applications. Conversely, code injection enables adversaries to introduce malicious code into a target process’s memory, modifying its execution flow. By targeting timing-related logic, attackers can distort an application’s perception of time. 

\vspace{-0.3cm}
\subsection{Attack Strategies}
Time manipulation attacks can be categorized into strategies based on how they distort an application's perception of time. We examine three primary strategies: constant attacks, incremental attacks, and randomized delay attacks. \textbf{Constant attacks} introduce a fixed offset while preserving the monotonic progression of time, resulting in a systematic deviation. \textbf{Incremental attacks} progressively modify time offsets, leading to a perceived time acceleration. \textbf{Randomized attacks} introduce stochastic fluctuations in time measurements, disrupting deterministic timekeeping.

\vspace{-0.3cm}
\subsection{Threat Model}
To reason about the practicality and viability of these attacks, we use a standard threat model used in prior timing attack work \cite{timeguard}. The\textbf{ threat model} assumes a privileged adversary capable of executing code, accessing system resources, and hijacking exception handlers. To gather data on system call usage, the attacker can profile the target program and take control of infrequently used global variables in the untrusted kernel. However, the attacker cannot maintain real-time shadow data structures such as shadow page tables or decompile binaries due to synchronization limitations.

\subsection{Privilege Escalation Attacks}
\begin{figure}[htbp]
    \centering
    \includegraphics[width=0.9\columnwidth]{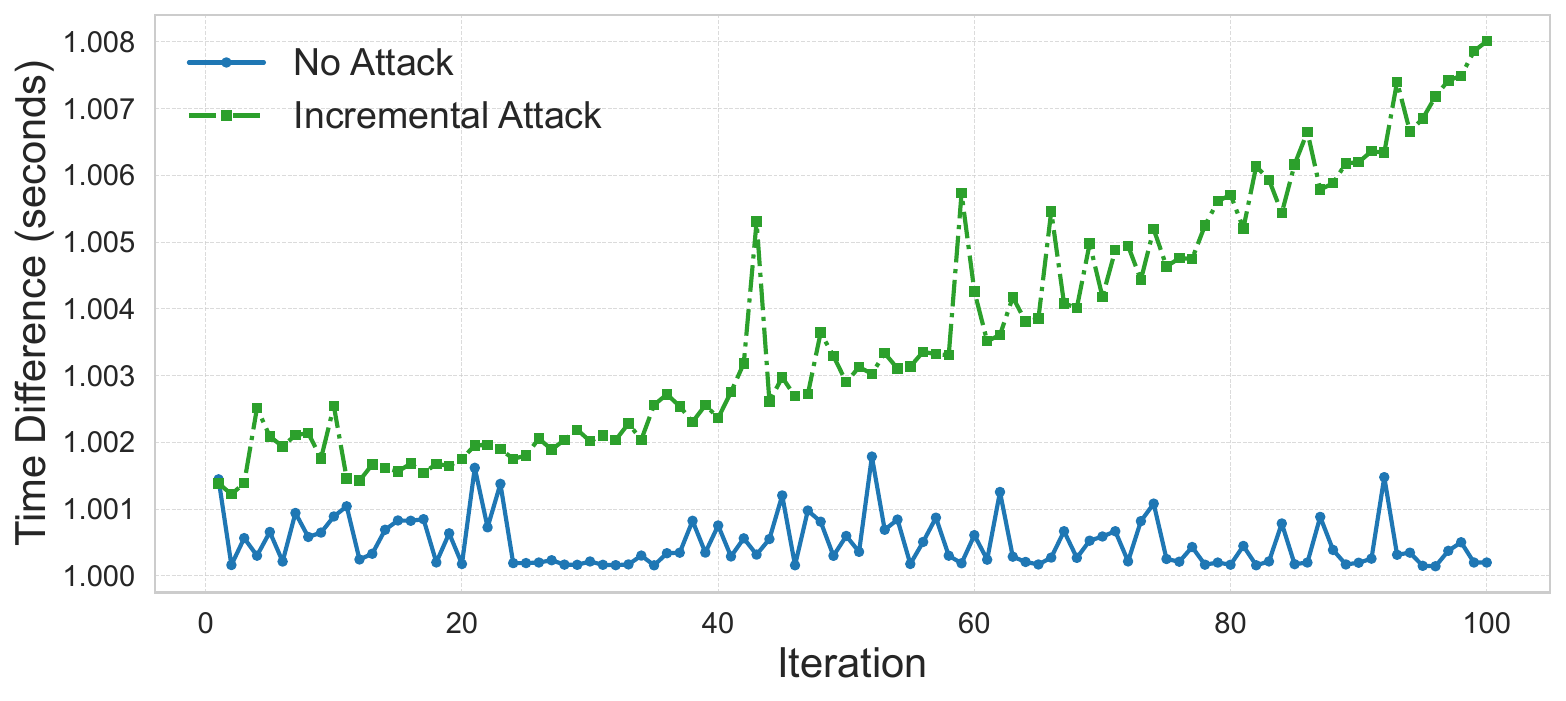}
    \vspace{-0.4cm}
    \caption{Time differences between iterations under normal and incremental attack conditions.}
    \vspace{-0.5cm}
    \label{fig:time_attack}
\end{figure}

We analyze two privilege escalation techniques: dynamic library injection and system call modification. \textbf{Dynamic library injection} exploits the dynamic linking mechanism of modern operating systems to modify an application’s runtime behavior. By loading malicious libraries into a process’s address space, attackers can intercept, manipulate, or override standard library functions. We simulate this attack using the \texttt{LD\_PRELOAD} environment variable on Linux. This vector is highly versatile, requiring no modification to the application binary, leaving no persistent system changes, and introducing minimal overhead. Using this method, we injected a custom library that intercepts an application’s calls to \texttt{clock\_gettime()} and returns manipulated values. The attack was tested under multiple conditions, including constant and incremental shift attacks. Figure~\ref{fig:time_attack} illustrates an attack scenario where the delay starts at 100 ms and increases by 1\% with each successive call.

\textbf{System call modification} enables attackers to alter the behavior of critical kernel functions, including those responsible for timekeeping. In our experiments, we used \texttt{bpftrace} kernel probes to attach adversarial code to timing-related system calls. These probes intercepted each syscall invocation, allowing us to monitor and modify the returned time values.

\subsection{Code Injection Attacks}
The attack was conducted in a series of stages. First, we attached to the target process using \texttt{ptrace}, gaining control over its execution. Next, we performed dynamic analysis to locate timing-related functions and variables. Once critical areas were identified, we injected malicious code by redirecting function calls to custom implementations. Finally, the injected code modified time values returned to the application, distorting its perception of time.

\section{Evaluation}
\begin{figure}[h!]
\centering
\includegraphics[width=0.9\columnwidth,keepaspectratio]{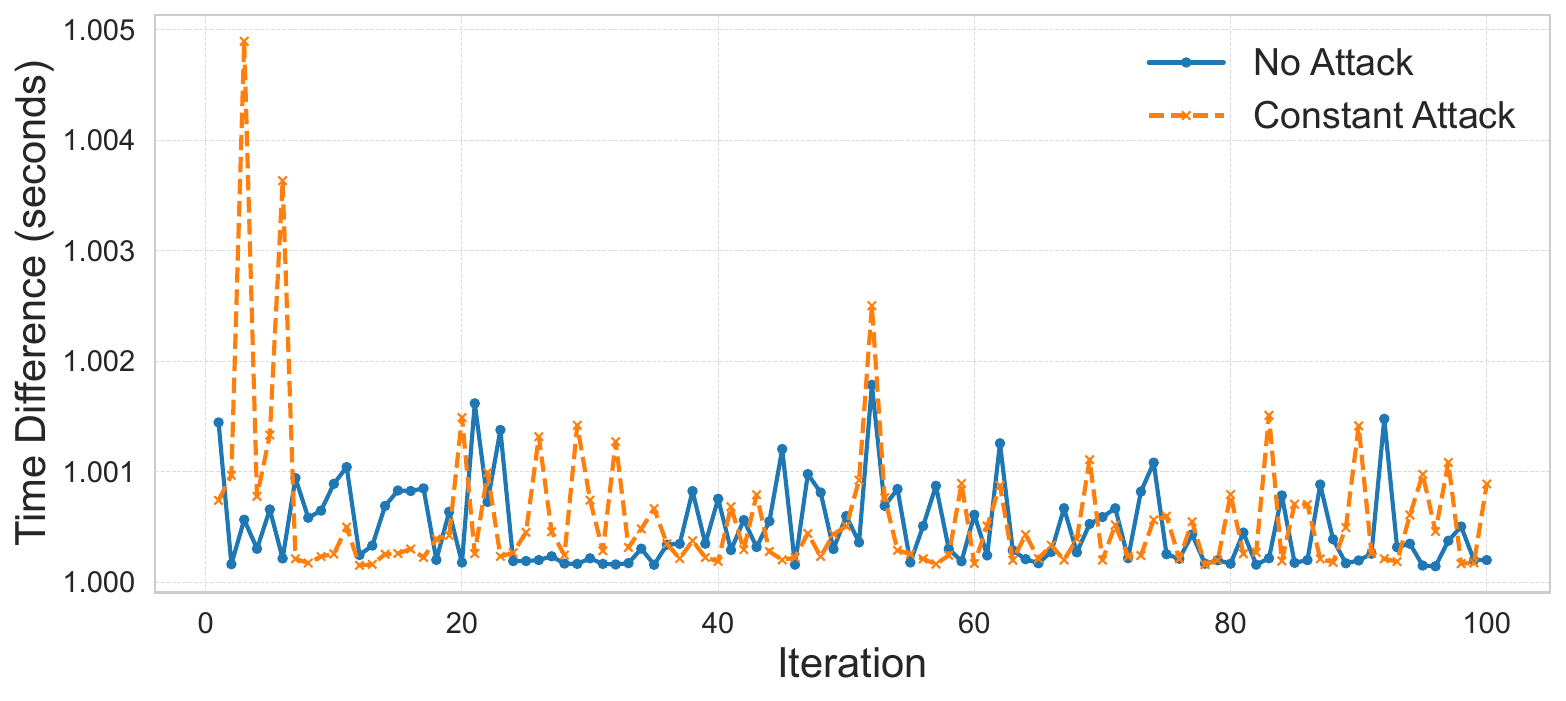}
\vspace{-0.5cm}
\caption{Overhead under system call modification attack.}
\vspace{-0.3cm}
\label{fig:attack_ov1}
\end{figure}

We initially categorize the overhead introduced by timing attacks on user applications. Figure~\ref{fig:attack_ov1} illustrates the overhead of a syscall modification attack introduced on a sample application. The application was tasked with fetching the time at $1$-second intervals. Under attack conditions, we observed an average delay of $1 ms$ in the system's time retrieval. A delay this small can quickly go undetected due to unpredictable system noise. 

\begin{figure}[htbp]
    \centering
    \includegraphics[width=\columnwidth]{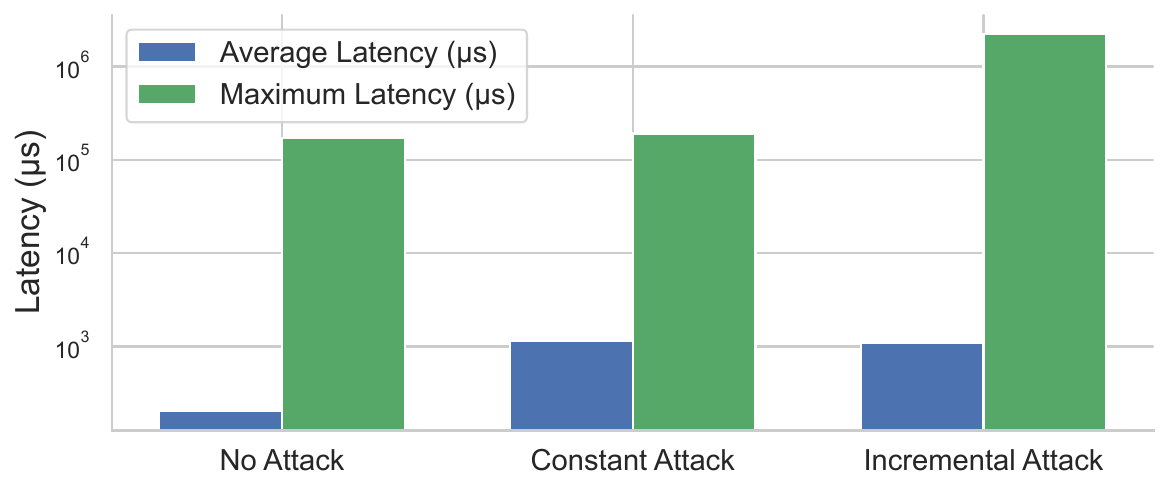}
    \vspace{-0.8cm}
    \caption{Impact on average and maximum latency from timing attacks on the \texttt{cyclictest} benchmark.}
    \vspace{-0.6cm}
    \label{fig:latency}
\end{figure}

We run \texttt{cyclictest} benchmark under normal and attack conditions to evaluate the efficacy and stealthiness of timing attacks. Figure \ref{fig:latency} shows one such case, where we test the system under no attack, constant attack, and incremental attack. We see that the latency increases significantly under constant and incremental attacks, with average latency rising roughly $6\times$. The significant increase in maximum latency under incremental attack shows severe system degradation over time. This implies that a constant attack consistently impacts performance, and an incremental attack progressively worsens the system behavior.  

\section{Conclusion}
Our study demonstrates the critical security risks of time manipulation attacks exploiting kernel vulnerabilities. Through privilege escalation and code injection techniques, adversaries can distort an application's perception of time, leading to severe disruptions in time-sensitive systems. Our experimental evaluation highlights the stealthy nature of these attacks and their potential to compromise critical applications.
Our findings underscore the urgency of treating time security as a fundamental aspect of system integrity. As adversaries continue to refine kernel exploitation techniques, further research is needed to develop lightweight yet effective solutions that safeguard time-sensitive applications against malicious distortions. Addressing these vulnerabilities will be crucial in maintaining the reliability and security of modern computing systems.

\bibliographystyle{ACM-Reference-Format}
\bibliography{references}

\end{document}